\documentclass[twocolumn]{aastex631}

\newcommand{\kms}{{\rm km~s^{-1}}}

\newcommand{\oiii}{[\textrm{O}~\textsc{iii}]}

\newcommand{\ci}{[\textrm{C}~\textsc{i}]}

\newcommand{\co}{CO(7-6)}

\newcommand{\targ}{MRG-M2129}
\newcommand{\simgt}{\,\rlap{\lower 3.5 pt \hbox{$\mathchar \sim$}} \raise
1pt \hbox {$>$}\,}
\newcommand{\simlt}{\,\rlap{\lower 3.5 pt \hbox{$\mathchar \sim$}} \raise
1pt \hbox {$<$}\,}
\newcommand{\Msun}{M_{\odot}}

\newcommand{\logm}{\log M_*/\Msun}

\newcommand{\ha}{${\rm H\alpha}$}
\newcommand{\hb}{${\rm H\beta}$}

\newcommand{\spit}{{Spitzer}}

\newcommand{\Z}{{\rm [Z/H]}}

\newcommand{\pix}{{\tt piXedif}}

\def\doff{62} 
\def\doffkpc{169} 

\def\gdr{$\delta_{\rm GDR}$}

\submitjournal{ApJ}

\shorttitle{Compact dust emission in a gravitationally lensed massive quiescent galaxy at $z=2.15$}
\shortauthors{Morishita et al.}
\graphicspath{{./}{figures/}}

\begin{document}

\title{
Compact dust emission in a gravitationally lensed massive quiescent galaxy at $z=2.15$ revealed in $\sim130$\,pc-resolution observations by ALMA
}

\correspondingauthor{Takahiro Morishita}
\email{takahiro@ipac.caltech.edu}

\author[0000-0002-8512-1404]{T. Morishita}
\affiliation{IPAC, California Institute of Technology, MC 314-6, 1200 E. California Boulevard, Pasadena, CA 91125, USA}

\author[0000-0002-5258-8761]{Abdurro'uf}
\affiliation{Institute of Astronomy and Astrophysics, Academia Sinica, 11F of AS/NTU Astronomy-Mathematics Building, No.1, Sec. 4, Roosevelt Road, Taipei 10617, Taiwan}

\author[0000-0002-4189-8297]{H. Hirashita}
\affiliation{Institute of Astronomy and Astrophysics, Academia Sinica, 11F of AS/NTU Astronomy-Mathematics Building, No.1, Sec. 4, Roosevelt Road, Taipei 10617, Taiwan}

\author[0000-0001-7769-8660]{A. B. Newman}
\affiliation{The Observatories of the Carnegie Institution for Science, Pasadena, CA, USA}

\author[0000-0001-9935-6047]{M. Stiavelli}
\affiliation{Space Telescope Science Institute, 3700 San Martin Drive, Baltimore, MD 21218, USA}

\author[0000-0003-1564-3802]{M. Chiaberge}
\affiliation{Space Telescope Science Institute for the European Space Agency (ESA), ESA Office, 3700 San Martin Drive, Baltimore, MD 21218, USA}
\affiliation{The William H. Miller III Department of Physics and Astronomy, Johns Hopkins University, Baltimore, MD 21218, USA}



\begin{abstract}
We present new observations of \targ, a quiescent galaxy at $z=2.15$ with the Atacama Large Millimeter/submillimeter Array (ALMA). With the combination of the gravitational lensing effect by the foreground galaxy cluster and the angular resolution provided by ALMA, our data reveal 1.2\,mm continuum emission at $\sim130$\,pc angular resolution. Compact dust continuum is detected at $7.9\,\sigma$ in the target but displaced from its stellar peak position by $\doff\pm{38}$\,mas, or $\sim \doffkpc \pm 105$\,pc in the source plane. We find considerably high dust-to-stellar mass ratio, $4\times10^{-4}$. From non-detection of the [\textrm{C}~\textsc{i}]\,$^3{\rm P}_2 \rightarrow\,^3{\rm P}_1$ line, we derive $3\,\sigma$ upper limits on the molecular gas-to-dust mass ratio $\delta_{\rm GDR}<60$ and the molecular gas-to-stellar mass ratio $f_{\rm H2}<2.3\%$. The derived $\delta_{\rm GDR}$ is $\simgt2\times$ smaller than the typical value assumed for quiescent galaxies in the literature. Our study supports that there exists a broad range of $\delta_{\rm GDR}$ and urges submillimeter follow-up observations of quenching/recently quenched galaxies at similar redshifts. Based on the inferred low $\delta_{\rm GDR}$ and other observed properties, we argue that the central black hole is still active and regulates star formation in the system. Our study exhibits a rare case of a gravitationally lensed type 2 QSO harbored by a quiescent galaxy.
\end{abstract}



\section{Introduction} \label{sec:intro}
Quenching, or shutdown of star formation activity, is one of the key evolutionary phenomena of galaxies toward the local massive populations such as giant elliptical galaxies. Synergies of wide-field surveys and subsequent spectroscopic followups with large telescopes have allowed us to see the emergence of a massive ($\logm\simgt11$, where $M_*$ represents stellar mass), quiescent galaxy population as early as redshift $z\sim4$, when the universe is only a few billion years old \citep{marsan15,glazebrook17,schreiber18b,tanaka19,valentino20b}. For such massive galaxies to exist at such an early time, they not only need to form a significant amount of stars, but also stop their star formation activity very rapidly. Extremely intense star formation is observed in sub-millimeter (submm) galaxies and quasars in earlier epochs \citep[e.g.,][]{casey19,spilker20}, some of which are presumably the progenitors of massive quiescent galaxies seen at lower redshifts \citep{toft14,straatman14}. However, it is still an open question how galaxies stopped such extreme star formation, especially in the peak epoch of cosmic star formation \citep[][]{madau14b}. Understanding the corresponding physical mechanism(s) is critical for theoretical studies, as the predicted number densities of quiescent galaxies at those redshifts span an order of magnitude among studies \citep[e.g.,][]{valentino20b}.

To advance our understanding of galaxy quenching, one powerful key probe is inter-stellar medium (ISM), such as molecular gas and dust. Observations revealed only a small amount of molecular gas in nearby early-type galaxies --- $\simlt$ a few percent of the total stellar mass \citep[][]{leroy08,saintonge11}, which is an order of $\sim 2$ magnitude lower than typical star-forming galaxies \citep{boselli14,davis14}, suggesting that their low star-formation activity is largely due to the lack of cold gas. 

However, investigation of ISM in high-$z$ quiescent galaxies remains challenging due to, by definition, their inactive nature, except for a few cases \citep[e.g.,][]{schreiber18,williams21}. Still, recent deep submm observations have made significant progress via the detection of far-infrared emission from thermal dust, as an indirect inference of gas abundance from its tightly correlated origin \citep[e.g.,][]{remy-ruyer14}. Stacking analysis of FIR emission revealed a significant amount of gas ($\sim5$-$10\,\%$ of the stellar mass) in quiescent galaxies up to $z\sim2$ \citep[][]{gobat18,magdis21}. Some attempts have revealed gas reservoirs in individual galaxies \citep{whitaker21}, if not all of them \cite[e.g.,][]{spilker18,bezanson19,morishita21,bezanson22}. These observations suggest that gas depletion may not be the only cause for quenching at high redshifts, which is in line with early studies of lower-$z$ galaxies \citep[e.g.,][]{davis14,alatalo15,suess17}.

Further investigation of the spatial distribution of dust in those quiescent galaxies is likely the next key step to our understanding of how they ceased star formation and remain quiescent. To reach the characteristic scale of star formation and active galactic nucleus (AGN) activities, gravitational magnification is the only chance for us, even with the exquisite sensitivity and angular resolution provided by current facilities. 

In this paper, we present our new observations with the Atacama Large Millimeter/submillimeter Array (ALMA) of a massive quiescent galaxy, \targ\ at $z=2.15$. 
{
\targ\ is located in the sightline of a massive galaxy cluster, MACS2129-0741, at $z = 0.57$ \citep{ebeling07}. For its strong lens magnification by the foreground galaxy cluster, \targ\ is an ideal target to study its stellar and ISM properties in detail. \citet{geier13} conducted near-IR spectroscopic observations and found that \targ\ consists of old ($\sim1.7$\,Gyr) stellar populations. In addition, \targ\ is known for several other interesting properties; \citet{toft17} revisited the spectroscopic data of \citet{geier13} and found ordered rotation in its stellar disk; \citet{newman18} conducted follow-up spectroscopic observations and reported that its core is classified as Seyfert via emission-line diagnostics (but see Sec.~\ref{sec:dust_prop} where we find it likely being a type 2 quasar); \citet{whitaker21} revealed a significant amount of dust from their ALMA Band 6 observations; \citet{akhshik22} analyzed HST's NIR grism data and inferred relatively fast timescale of quenching in its core ($\sim0.3$\,Gyr) among their 8 quiescent galaxies at $z\sim2$.}
By taking advantage of the rich data sets collected by those early pioneering studies, we aim to advance our understanding of quenching in \targ. With the spatial resolution provided by ALMA and strong lens magnification by the foreground cluster, our observations aim to reveal the distribution of ISM in \targ\ at an unprecedented angular scale, $\sim130$\,pc, for a quiescent galaxy at this redshift. 

Throughout this paper, we adopt the AB magnitude system \citep{oke83,fukugita96}, cosmological parameters of $\Omega_m=0.3$, $\Omega_\Lambda=0.7$, $H_0=70\,\kms\, {\rm Mpc}^{-1}$, and the \citet{2003Chabrier} initial mass function.

\begin{figure*}
\centering
	\includegraphics[width=0.95\textwidth]{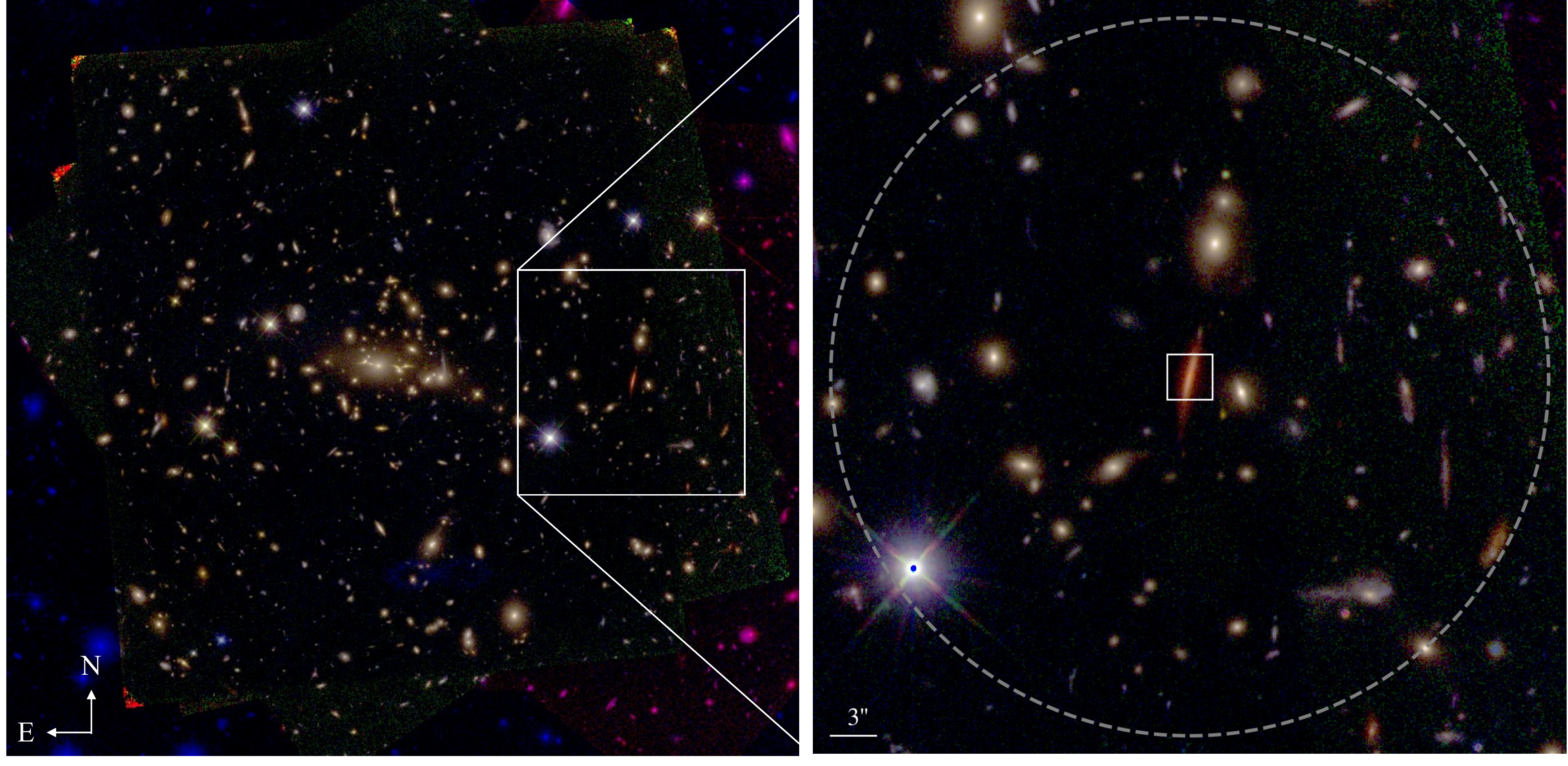}
	\caption{(Left:)
	HST's RGB composite image of MACS~2129-0741, a cluster of galaxies at $z=0.57$. The target of our ALMA observations, \targ, is located in the center of the white square box. 
	Right: Zoom-in image of \targ\ ($50''\times50''$). The FoV of our ALMA Band 6 observations is shown by the central square box of $3''\times3''$ (also see Figure~\ref{fig:alma}). The FoV of one of the previous Band~6 programs (2018.1.00276.S) is shown with a gray circle ($r\sim24''$) for comparison.
	}
\label{fig:mosaic}
\end{figure*}

\section{Data} \label{sec:data}
\subsection{ALMA Cycle 8 observations} \label{sec:alma}
Our interferometric observations with Band 6 were executed during ALMA Cycle 8 (2021.1.00847.S, PI T. Morishita) over three separate blocks (October 11, 14, and 15 2021), with the antenna configuration of C43-8. The pointing of our observations is shown in Fig.~\ref{fig:mosaic}. The choice of Band 6 was made based on two considerations --- 1.~robust dust estimate by detecting continuum flux at $\sim1.2$\,mm and 2.~spectral coverage of the \ci\,$^3{\rm P}\!_2 \rightarrow\,^3{\rm P}\!_1$ line (rest-frame wavelength $370.42\,\mu$m; hereafter \ci). The observed frequency range is an efficient window for ALMA, without being severely affected by atmospheric absorption, and thus provides a sweet spot for thermal dust emission originating from extra-galactic sources at $z\simgt1$. \ci\ is a fine structure line and known as an excellent tracer of cold gas in extra-galactic systems, especially at redshifts where the commonly used CO(J=1-0) is not available \citep{weiss03,walter11}. In addition, since \ci\ is optically thin, it is suited for a wide range of star-forming properties and redshifts \citep{papadopoulos04,walter11,alaghband-zadeh13}. 

Each of the four spectral windows was configured with a bandwidth of 0.938\,GHz and 1920 frequency channels. We placed one of the spectral windows at $257.1$\,GHz, targeting the redshifted \ci\ line. We centered another spectral window at 256.5\,GHz, to cover the \co\ line. The other two spectral windows were placed at 243\,GHz and 246\,GHz so that we can robustly estimate the continuum level without being contaminated by any bright emission line. The astrometric calibration was done with two quasars (J2258-2758, J2130-0927), and the flux calibration by targeting J2258-2758. The observations were completed without raising severe flags, resulting in the astrometric uncertainty of $\sim 5.5\%$ of the beam size. The on-source exposure time was 139\,minutes. 

The data are reduced by using the public version of CASA (v.6.4.4). To inspect emission lines, we first subtract the continuum in the uv-space by using the {\tt uvcontsub} task of CASA after manually flagging visibility data. For robust continuum estimate, we exclude the spectral window that covers the \ci\ line but still use the one for the CO line, because the contribution from the line turned out to be negligible (see below). The estimated continuum flux is extrapolated to the spectral window of \ci\ and subtracted from it. We examined the results with two continuum models with a different polynomial order, 0 (constant) and 1, and found that the latter presents smaller residual fluxes in the two spectral windows located at line-free frequency ranges. We then run the {\tt tclean} task of CASA on the continuum subtracted visibility map, with a velocity resolution element $\sim50$\,km\,/\,s. While various sets of parameters are examined, we do not confidently detect the \ci\ nor \co\ lines in any case (Appendix~A).  

For continuum imaging, we run the {\tt tclean} task of CASA with two different setups. One opts for high-resolution imaging, {with briggs weighting}, the robust parameter set to 0.5, and the final pixel scale to $0.\!''01$. The final beam size is $0.\!''077\times0.\!''060$ with beam angle of $-70^{\circ}$. The image is used to infer flux distribution but not for absolute flux measure. For flux measurement, we create a continuum image with a low-resolution setup, with natural weighting, where the robust parameter is set to 2.0, and the pixel scale to $0.\!''1$, and taper size of $0.\!''3$, where the beam size is $0.\!''145\times0.\!''136$. Since no confident lines are identified over the frequency range, we use the whole visibility data to make a continuum image to accommodate higher signal-to-noise ratios (but see Section~\ref{sec:gasmass}).

To supplement our spectral energy distribution analyses in Sec.~\ref{sec:gasmass}, we retrieve archival data taken by ALMA with Band~3 (2016.1.01591.S, PI J. Zabl) and Band~6 (2018.1.00035.L, PI K. Kohno; 2018.1.00276.S, PI K. Whitaker). The continuum images of each program were retrieved through the ALMA Science Archive. One of the Band~6 observations shows a clear but unresolved detection of a source \citep{whitaker21}. We do not detect continuum emission in the other two data sets and thus use those to infer flux upper limits.

\begin{figure*}
\centering
	\includegraphics[width=0.99\textwidth]{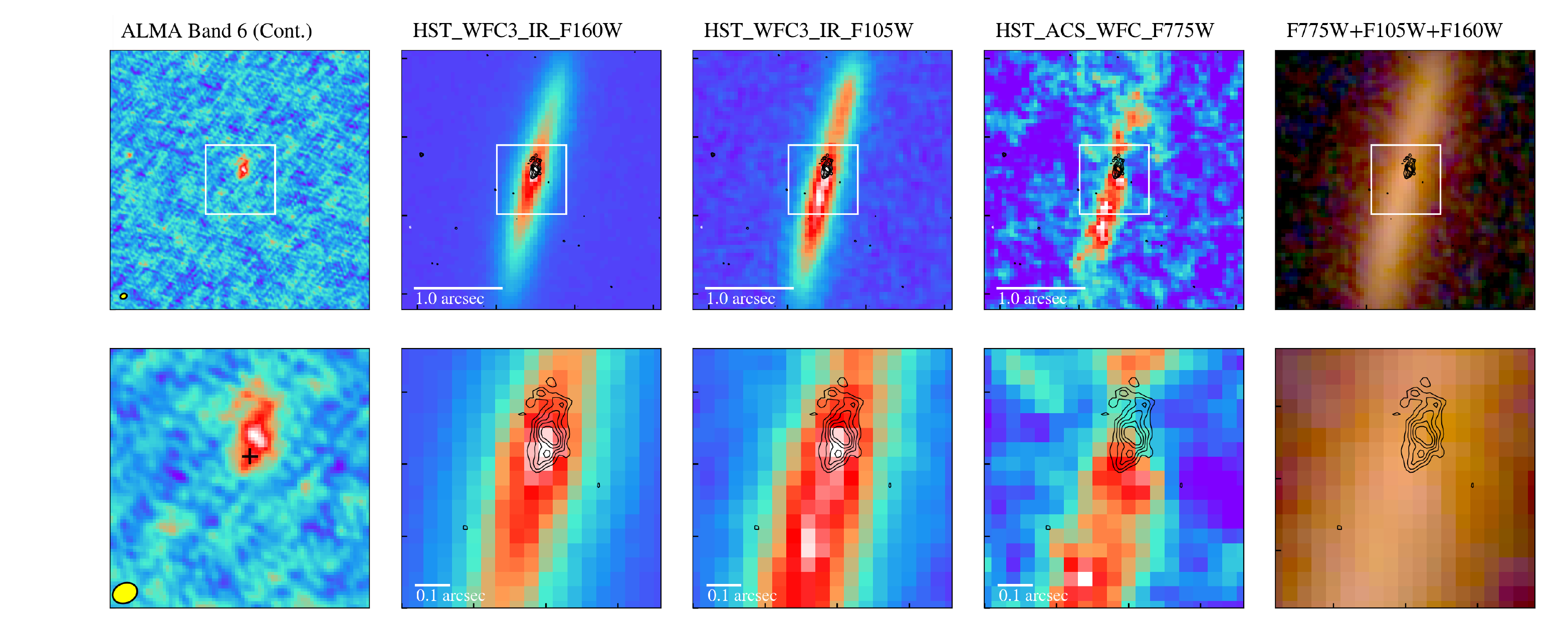}
	\caption{Top: Multi-band images of \targ\ in a $3''\times 3''$ cutout (same size as the white square in Fig.~\ref{fig:mosaic}) --- ALMA 1.2\,mm continuum created with the high-resolution setup (Sec.~\ref{sec:alma}), HST F160W, F105W, F775W, and pseudo RGB images, from left to right. The continuum flux of ALMA is overlaid in each of the HST images, where the contour represents 3, 4, 5, 6, and 7\,$\sigma$. The beam size is shown at the bottom left of the ALMA image. Bottom: zoomed-in version of the images in the top row in a box of $1''\times1''$, corresponding to the white square box in the top panel. The flux centroid position measured in the F160W image (Sec.~\ref{sec:morp}) is shown in the 1.2\,mm image (cross symbol).
	}
\label{fig:alma}
\end{figure*}

\subsection{Supplemental HST and Spitzer images} \label{sec:hst}
We collect the optical to near-infrared images taken in a series of past campaigns, including  HST's CLASH \citep{postman12}, GLASS \citep{schmidt14,treu15}, REQUIEM, \citep{akhshik21} and \spit\ SURFSUP survey \citep{bradac14,huang16}. 

For HST images, the raw data are acquired through the MAST archive and processed by an imaging reduction pipeline, {\tt borgpipe} \citep{morishita21b}. One of the advantageous points of reprocessing the data is that the new pipeline enables astrometric alignment to the latest GAIA WCS frame. This is critical for us regarding the comparison with a dataset taken by ALMA (see Sec~\ref{sec:morp}). During the pipeline processing, images are aligned to the GAIA DR2 world coordinate system (WCS) frame by using the stars available within the FoV of the image. The final RMSs in the astrometry are $0.\!''035$ and $0.\!''040$ for RA and Dec, respectively. The aligned images are then processed with the astrodrizzle package, with the final pixel scale set to $0.\!''045$.

The flux error map of each filter is scaled to the flux RMS measured in the empty region of the science image, to take account of correlation noise caused by drizzling \citep[e.g.,][]{calvi16,morishita18b}.

Spitzer IRAC (channels 1, 2, and 4) and MIPS (ch1) data are retrieved through the Spitzer Legacy Archive\footnote{\doi{10.26131/irsa408}.}. We use a public python package {\tt golfir}\footnote{\url{https://github.com/gbrammer/golfir}.} to reduce the raw data. After background subtraction and alignment of each image in the pbc format, we drizzle and combine images by setting the final pixel scale to $0.\!''54$ and the pixel fraction to 0.4. Combined images are then aligned to the HST F160W image processed above.

For spectral energy distribution analysis of \targ\ (Sec.~\ref{sec:sed_global}), we extract photometric fluxes. For the HST images, we use ISOPHOT\_FLUX measured by {\tt SExtractor} \citep{bertin96}. The segment image is defined with the F125W+F140W+F160W-stacking image. For the \spit\ IRAC images, we model the observed flux distribution of each channel by using the source morphology modeled in F160W as a prior, in the same way as presented in \citet{morishita20}. While the model has been done in the lens plane assuming a one-component S\'ersic light profile, the results are characterized by reasonable residual that is within the uncertainty level of the original image. For the MIPS image, we do not detect emission at the source position and thus place a flux upper limit by calculating the RMS measured in the nearby empty region with $r=7''$ aperture.

\begin{deluxetable}{lcc}
\tabletypesize{\footnotesize}
\tablecolumns{3}
\tablewidth{0pt} 
\tablecaption{Flux centroids of \targ.}
\tablehead{
\colhead{Images} & \colhead{R.A.} & \colhead{Decl.}
}
\startdata
HST F160W & 21:29:22.3446 & -07:41:30.966\\
ALMA 1.2\,mm (This work) & 21:29:22.3433 & -07:41:30.907\\
\enddata
\tablecomments{
The associated systematic uncertainties are $\sim37.5$\,mas and $5.5$\,mas for the HST F160W and ALMA 1.2\,mm images, respectively (Sec.~\ref{sec:morp}).
}
\label{tab:centroid}
\end{deluxetable}

\begin{figure}
\centering
	\includegraphics[width=0.23\textwidth]{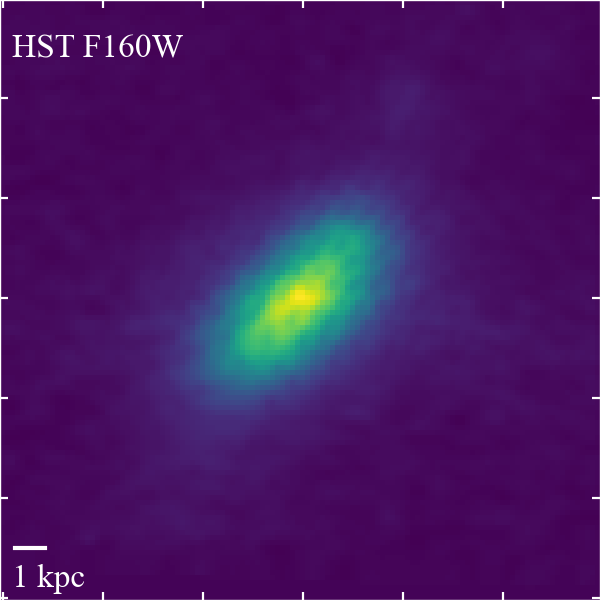}
	\includegraphics[width=0.23\textwidth]{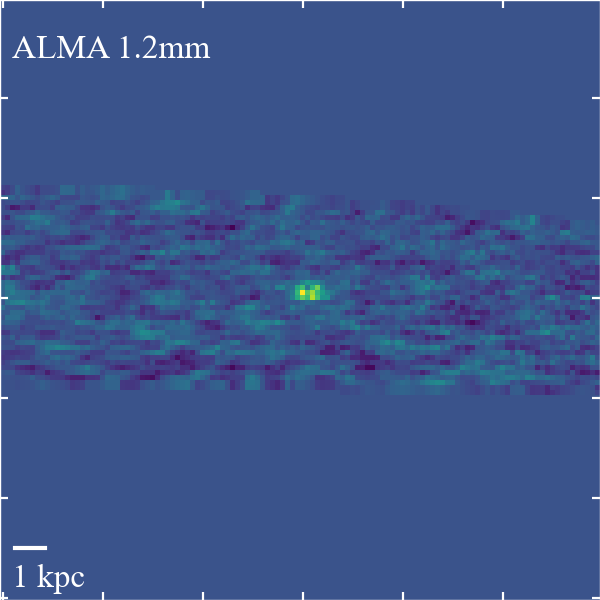}
	\caption{Source plane image of \targ, in HST F160W (left) and ALMA 1.2\,mm (right). The stamp size corresponds to $\sim20$\,kpc in the source plane. The HST (ALMA) image is smoothed with a Gaussian kernel of $\sigma=0.\!''2$ ($0.\!''04$) and shown using an arcsinh (linear) stretch.
	}
\label{fig:image_sp}
\end{figure}

\section{Results}
\subsection{Morphology in the 1.2\,mm continuum}\label{sec:morp}
In Fig.~\ref{fig:alma}, we show the high-resolution 1.2\,mm continuum image of \targ, which traces thermal emission from dust. With the spatial resolving power of ALMA combined with lens magnification by the foreground cluster (by a factor of $\mu\sim4.5$), the image reveals the morphology of the source at the resolution of $\sim60$\,pc in radius, an unprecedented resolution for a quiescent galaxy at $z\simgt0.1$. The continuum flux is detected at $7.9\,\sigma$ at the flux peak position. The apparent size of the emission is measured by the Two-D Fitting Tool provided by CASA, to be $\sim0.\!''22\pm0.03$ and $\sim0.\!''10\pm0.02$ along the major and minor axis, respectively, which is sufficiently resolved by our primary beam size. While the continuum morphology is elongated almost along the vertical axis (i.e. South-East to North-West), this is mostly due to the lens magnification effect along the same direction (see below). 

The HST images of the same region are shown in Fig.~\ref{fig:alma}. From the comparison of these images, we notice that the flux peak position in the 1.2\,mm image is located slightly off from the flux peak of the F160W image, which approximately traces the stellar mass distribution. We measure the flux centroid in each image by using {\tt centroid\_quadratic} of astropy (Table~\ref{tab:centroid}). The offset is measured as $\doff$\,mas.

It is noted that the two images have been independently aligned to the world coordinate system (WCS). The typical uncertainty of astrometry in ALMA, which was calibrated by using a quasar, is { calculated by $\theta_{\rm beam}/{\rm SN}/0.9$, where $\theta_{\rm beam}$ is the Full-Width Half Maximum synthesized beam size in arcseconds and SN is the source signal-to-noise ratio. This provides the pointing accuracy of our observations, $\sim10.6$\,mas.} There is no flag raised regarding the calibration. The HST images have been aligned to the GAIA DR2 by using a package {\tt tweakreg}. For the alignment, we identified seven stars in the field by cross-matching with the GAIA DR2 catalog. The final standard deviation of the residual shifts is $37.5$\,mas. After taking those systematic uncertainties into account ($\Delta=39.0$\,mas), the measured offset is $\sim1.6\,\sigma$ significant and thus we consider this as tentative.

To infer the physical distance of the observed offset, we correct lens magnification in both images. We use the lens magnification model adopted in \citet{newman18b}, which was originally presented in \citet{monna17}. The source plane images are shown in Fig.~\ref{fig:image_sp}. The physical distance between the two centroids is measured in the source image and inferred to be $\sim \doffkpc\pm{105}$\,pc. 

We fit a two-dimensional Gaussian to the source plane image and measure the physical size of the 1.2\,mm emission to be $\sim0.44$\,kpc and $0.22$\,kpc along the major and minor axes, respectively, characterizing its compact morphology.

In Fig.~\ref{fig:alma}, we also show the HST/ACS F775W image, which traces young stellar populations (at rest-frame $\sim0.25\,\mu$m). We find that the F775W flux distribution does not overlap with the ALMA continuum. This suggests that short-wavelength emission is attenuated at the position of the dust continuum peak, which also assures that the detected dust continuum is likely associated with the system, not a chance superposition.

With the low-resolution image, we measure the total 1.2\,mm flux of $0.19\pm0.03$\,mJy by using a circular aperture of $r=0.\!''14$. The flux uncertainty is estimated by the RMS of fluxes in a set of $r=0.\!''14$ apertures randomly placed around (but not on) the emission within $0.\!''26$ from the peak position. The measured flux is slightly larger than the previous Band~6 measurement in \citet{whitaker21}, $0.16\pm0.02$\,mJy (K. Whitaker, priv. comm.), but this reflects the fact that they were taken at different frequencies (1.2 and 1.3\,mm). The measured fluxes are in reasonable agreement given the spectral curve of dust emission (see Sec.~\ref{sec:sed_global}). 

It is worth noting that the FoV of our observations is much smaller than the one in \citet{whitaker21}, where the maximum recoverable scale is $\sim 9.\!''5$ with the beam size of $\sim1''$ (Fig.~\ref{fig:mosaic}). The agreement in the measured fluxes of the two observations implies that the dust emission seen in the previous study mostly originates in the compact region revealed here.

\begin{figure}
\centering
	\includegraphics[width=0.5\textwidth]{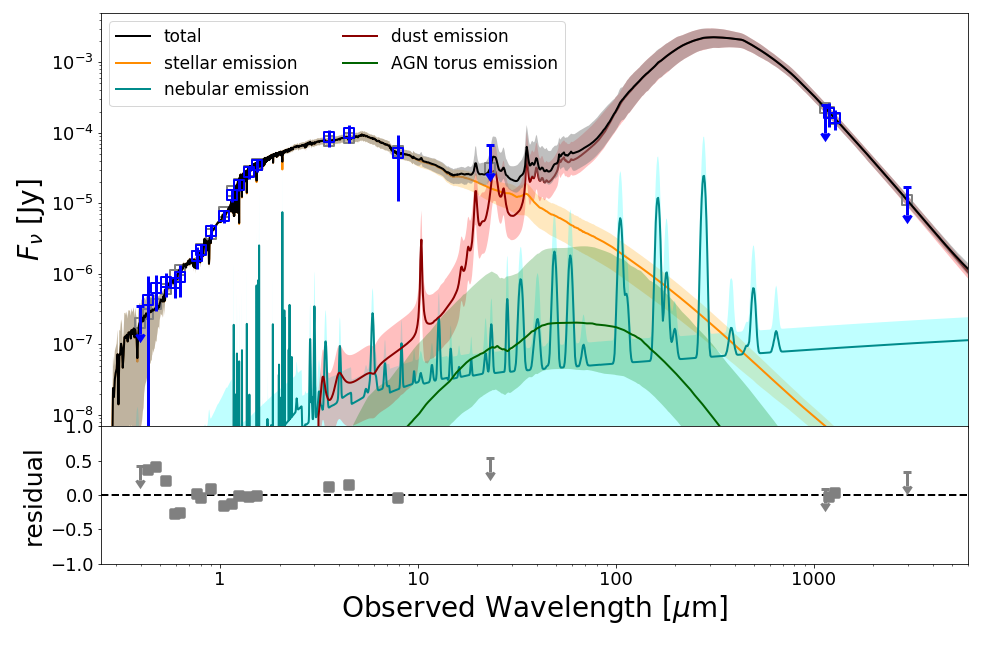}
	\includegraphics[width=0.5\textwidth]{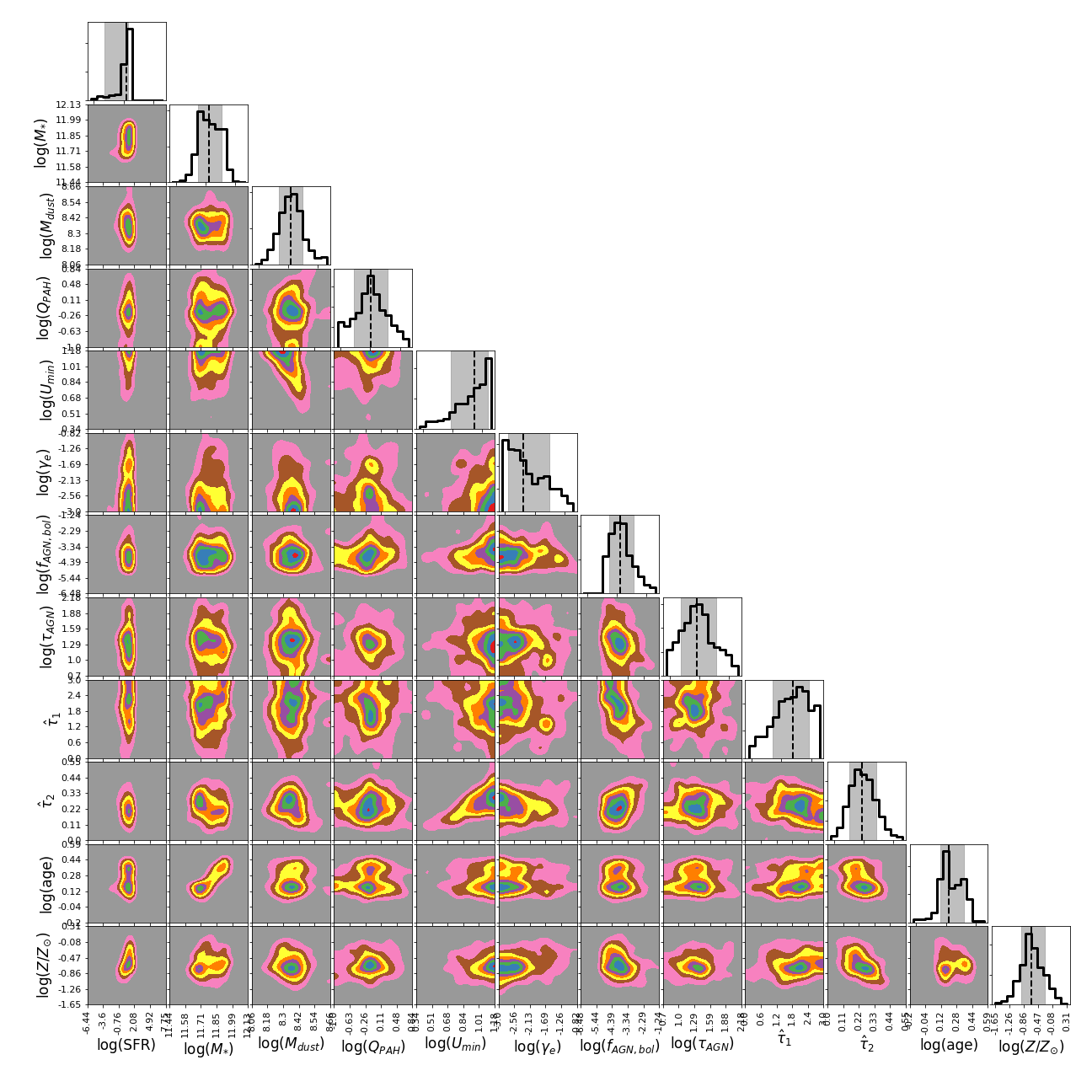}
	\caption{
	(Top)
	Spectral energy distribution (SED) of integrated fluxes of \targ. The best-fit model template is shown in a solid black line. Each component of the fitting template is shown with a different color scheme as in the label. Data points are shown with $2\,\sigma$ errors (blue symbols). $2\,\sigma$ upper limits are shown for non-detection. Normalized residual, $(\text{data}-\text{model})/\text{data}$, is shown in the bottom panel with the same symbols as in the top panel.
	{(Bottom) Posterior distributions of the parameters used in our SED fitting analysis.}
}
\label{fig:sed_global}
\end{figure}

\begin{deluxetable}{lr}
\tabletypesize{\footnotesize}
\tablecolumns{2}
\tablewidth{0pt} 
\tablecaption{Best-fit parameters of SED modeling of MRG-M2129 \label{tab:fit_global_sed}}
\tablehead{
\colhead{Parameter} & \colhead{Best-fit value}
}\label{tab:sed}
\decimals
\startdata
$\log(M_{*}[M_{\odot}]\times\mu)$ & $11.78_{-0.10}^{+0.13}$ \\
$\log(M_{\rm d}[M_{\odot}]\times\mu)$ & $8.36_{-0.10}^{+0.10}$ \\
$\log(\text{SFR}[M_{\odot}\text{yr}^{-1}]\times\mu)$ & $0.64_{-5.40}^{+0.31}$ \\
$\log(\text{sSFR}[\text{yr}^{-1}])$ & $-11.20_{-5.33}^{+0.36}$ \\
$\log(\rm{age[Gyr]})$ & $0.20_{-0.10}^{+0.17}$ \\
$\log(Z[Z_{\odot}])$ & $-0.68_{-0.30}^{+0.37}$ \\
$T_{\rm d}$[K] & $26.77_{-1.74}^{+1.53}$ \\
$\log(L_{\rm IR}[L_{\odot}]\times\mu)$ & $11.51_{-0.17}^{+0.12}$ \\
$\log(U_{\rm min})$ & $0.98_{-0.26}^{+0.15}$\\
$\log(\gamma_{e})$ & $-2.36_{-0.46}^{+0.79}$\\
$\log(Q_{\rm PAH})$ & $-0.15_{-0.43}^{+0.45}$ \\
$\log(f_{\rm AGN,bol})$ & $-3.87_{-0.76}^{+1.00}$ \\
$\log(\tau_{\rm AGN})$ & $1.35_{-0.34}^{+0.40}$\\
$\hat\tau_{1}$ & $1.86_{-0.86}^{+0.70}$ \\
$\hat\tau_{2}$ & $0.25_{-0.10}^{+0.10}$ \\
\enddata
\tablecomments{$\mu$: Magnification factor. 
$\hat\tau_{1}$: Dust optical depth of the birth cloud. $\hat\tau_{2}$: Dust optical depth of the diffuse ISM. 
$T_{\rm d}$: Dust temperature. 
$U_{\rm min}$: Minimum starlight intensity that illuminates the dust. 
$\gamma_{e}$: Relative fraction of dust heated at a radiation field strength of $U_{\rm min}$ and dust heated at $U_{\rm min}<U\leq U_{\rm max}$. $Q_{\rm PAH}$: Percentage fraction of total dust mass that is in the polycyclic aromatic hydrocarbons (PAHs). 
$L_{\rm IR}$: infrared bolometric luminosity summed over $8$--$1000\,\mu$m. 
$f_{\rm AGN,bol}$: fraction of the bolometric luminosity that is contributed by the AGN dusty torus emission. $\tau_{\rm AGN}$: Optical depth of the AGN dusty torus.
}
\end{deluxetable}

\subsection{Dust mass estimates}\label{sec:sed_global}

In the previous section, we presented a new flux measurement of \targ\ at 1.2\,mm. To validate the consistency with the previous measurement at 1.3\,mm \citep{whitaker21}, we here estimate dust mass by fitting a modified blackbody template to our 1.2\,mm flux data point. By following \citet{bianchi13}, dust mass is derived by;
\begin{equation}\label{eq:dust}
    M_{\rm d} = \frac{f_{\nu} D_{L}^2}{\kappa_{\rm abs} B_\nu(T_{\rm d})}
\end{equation}
for the observed flux, $f_{\nu}$, where $D_L$ is the luminosity distance to the source, $B_\nu(T_{\rm d})$ the Planck function at the temperature of $T_{\rm d}$. $\kappa_{\rm abs}$ is the grain absorption cross-section per unit mass, or called emissivity. We use the following formula from \citet{scoville14};
\begin{equation}
    \kappa_{\rm abs} = \kappa_{\rm abs}(\nu_0) \times ({\frac{\nu_0}{345}})^\beta.
\end{equation}
where $\kappa_{\rm abs}=0.0484\,{\rm m\,kg^{-2}}$ and $\nu_0$ is rest-frame frequency. We use $T_d=20$\,K and $\beta=1.8$, the same values used in \citet{whitaker21}. With those parameters, we obtain $M_{\rm d}\sim2\times10^8\,M_\odot$ (uncorrected for magnification), which is in excellent agreement with one by \citet[][$2\times10^8\,M_\odot$]{whitaker21}. This agreement indicates that the flux probed in our observations represents the whole flux revealed in the previous Band~6 observations that cover the entire stellar extent of \targ\ (Fig.~\ref{fig:mosaic}).

To characterize the FIR properties of \targ\ in further detail, we attempt to include multiple FIR data points collected from the literature. This supplement dataset spans from MIPS~$24\,\mu$m to ALMA Band~3 ($2.9$\,mm), as well as rest-frame UV-to-IR data (Sec.~\ref{sec:alma}). {While most FIR data points from literature are non-detection, flux upper limits of those data points still add constraints on FIR modeling.} Three UV-filter images (F225W, F275W, F336W) show positive fluxes but at $S/N\simlt2$. We visually inspected the images and only found slight fluctuation in background value around the position of \targ\ but no confident detection. We thus exclude those three data points from our SED fitting analyses.

For FIR modeling, it is critical to cover a wide range of wavelengths and take into account of energy balance between UV--optical and infrared. For this reason, we here use an SED modeling code, \pix\ \citep{abdurrouf21b}. \pix\ adopts Flexible Stellar Population Synthesis \citep[FSPS;][]{2009Conroy}, assuming the \citet{2003Chabrier} initial mass function, double power-law star formation history model, two-component dust attenuation law of \citet{2000Charlot}, \citet{2008Nenkova_a} AGN dusty torus emission model, and dust emission model of \citet{2007Draine}. We mainly adopt the same priors as those assumed in \citet{abdurrouf21b}, except for the following parameters: the power-law index of the dust attenuation model $n$=[$-1.5$, $0.4$], stellar age $\log(\text{age}[\text{Gyr}])$=[$-2.0$, $0.48$], and the exponential time scale of star formation $\log(\tau[\text{Gyr}])$=[$-1.0$, $1.5$]. Similarly to \citet{abdurrouf21b}, we assume uniform priors within those ranges. The interested reader is referred to \citet{abdurrouf21b} for the full description of the parameters. For the fitting method, we apply the Markov Chain Monte Carlo (MCMC) and use the number of walkers and step per walker of 100 and 1000, respectively.

{In Fig.~\ref{fig:sed_global}, we show the best-fit model along with the parameter distribution.} 
In Table~\ref{tab:sed}, we summarize the best-fit parameters. The best-fit result provides a slightly larger dust mass ($2.3\pm0.5\times10^{8}\,M_\odot$) and dust temperature ($26.8\pm1.6$\,K) than those derived or assumed in the analysis above. The derived temperature is in fact larger than the median value ($\sim21$--$23$\,K) of quiescent galaxies at $1<z<2.5$ \citep{gobat18,magdis21}, which implies that \targ\ is a young, recently quiescent galaxy compared to the average population used in those studies (see also Sec.~\ref{sec:dust_prop}). {Despite the inclusion of FIR data and different reduction of HST images, the derived stellar mass ($\logm\times\mu=11.78_{-0.10}^{+0.13}$) is broadly consistent with two previous measurements --- $11.62\pm0.05$ \citep[][]{newman18,whitaker21} and $11.8\pm0.20$ \citep[][]{toft17}. In what follows, we use the dust and stellar masses derived with \pix.}

We obtain dust-to-stellar mass ratio of $3.8_{-0.7}^{+1.0}\times10^{-4}$. This is considerably high among those of local early-type galaxies \citep[$\simlt 10^{-5}$; e.g.,][]{smith12} but similar to the median values of quiescent galaxies at $z\simgt1$ \citep[][{which adopt the \citealt{2007Draine} model for dust mass estimate}]{magdis21}. Despite the current star formation status of \targ\ (sSFR\,$\simlt10^{-11}$\,yr$^{-1}$), the observed high abundance of dust indicates that \targ\ experienced quenching in a relatively short time scale, possibly short enough before dust destruction processes become effective. {This conclusion is consistent with the star formation history inferred by \citet{akhshik22}, where they estimate the declining time scale of star formation to be $\sim0.3$\,Gyr in the central region.} We discuss this in Sec.~\ref{sec:disc}.

{It is noted that our use of integrated photometry for SED fitting implicitly assumes that the energy balance between UV-optical and FIR is conserved in the entire system. However, the assumption is not obviously validated for the case of \targ, where dust distribution is not co-spatial as the stellar emission. To investigate this, we repeat SED fitting process for the central region, by using only resolved images (i.e. HST and our Band~6 ALMA continuum). We extract fluxes across those images in a common region, defined by an $r<0.\!''5$ circular aperture centered on the dust continuum peak. The resulting dust and stellar masses in the central region are $3.0_{-2.0}^{+4.2}\times10^{8}\,M_\odot$ and  $3.1_{-1.1}^{+2.4}\times10^{11}\,M_\odot$, respectively. By adding the remaining stellar mass at the outer region ($\sim4\times10^{11}\,M_\odot$; derived only with the HST images), we obtain dust-to-stellar mass ratio of $\sim4.2\times10^{-4}$, which is consistent with our original estimate above within the uncertainty. The increase in the uncertainties of dust and stellar mass estimates here is attributed to the reduced number of data points.}

\subsection{Inference on molecular gas mass}\label{sec:gasmass}
In this subsection, we aim to place an upper limit on the molecular hydrogen mass in \targ. We follow a similar procedure presented in \citet{morishita21}. We first estimate the RMS of each low-resolution cube layer (ie. two-dimensional image at each frequency) by using the {\tt imstat} task of CASA. We then integrate the derived RMS values over the line width, $\Delta v = 2\sigma_v$, around the systemic redshift of \targ. We set $\sigma_v=261$\,km/s, which is the stellar velocity dispersion measured in \citet{newman18}. Since the dust emission is spatially resolved, we also scale the RMS values by integrating over the area of the dust detection, defined by an ellipse of $\sim0.\!''47 \times 0.\!''26$, corresponding to $5\,\sigma$ size obtained in two-dimensional gaussian fit. With these, we obtain a $3\,\sigma$ upper limit of the \ci\ velocity integrated flux $I_{\ci}<16.3$\,mJy~km/s. {We also repeated the same process in the negative cube and found a consistent flux limit.}

We then use the recipe described in \citet{man19,jiao19}, to convert the velocity integrated flux into molecular carbon mass $M_{\rm [CI]}$ by
\begin{equation}
    L'_{\rm [CI](2-1)} = 3.25\times 10^{3} \left[ \frac{D_L^2}{(1+z)}\right] \left( \frac{\nu_{\rm [CI](2-1),rest}}{\rm 100\,GHz}\right)^{-2} I_{\ci},
\end{equation}
\begin{equation}
    M_{\rm [CI]} = 4.566 \times 10^{-4} Q(T_{\rm ex}) \frac{1}{5} e^{{62.5/T_{\rm ex}}} L'_{\rm [CI](2-1)}
\end{equation}
under optically thin and local thermodynamical equilibrium assumptions. We set the \ci\ excitation temperature $T_{\rm ex}$ to $19.7$\,K, the median value of those of local star-forming galaxies in \citet{jiao19}. $Q_{\rm ex} = 1 + 3 e^{-T_1/T_{\rm ex}} + 5e^{-T_2 /T_{\rm ex}}$ is the \ci\ partition function, with $T_1=23.6$\,K and $T_2=62.5$\,K. 

Lastly, we convert the derived atomic carbon mass into molecular hydrogen mass by using a conversion factor, $X\ci/X{\rm [H_2]}$. This factor is known to have a wide range, from $\sim3\times10^{-5}$ \citep{weiss03,papadopoulos04} to $8.4\times10^{-5}$ \citep[][derived for SMGs and quasars]{walter11}, which may depend on redshifts, stellar populations, and/or star-forming phase \citep{jiao19}. Here, for the consistency of the $T_{\rm ex}$ value used above, we adopt the median value of the sample in \citet{jiao19}, {$2.5\times10^{-5}$}, which leads to molecular hydrogen mass $M_{\rm H_2}\simlt1.4\times10^{10}\,M_\odot$ ($3\,\sigma$; uncorrected for magnification). By taking the face value, this gives a molecular gas-to-dust mass ratio $\delta_{\rm GDR}\simlt60$, which is smaller than quiescent galaxy populations at $z\sim2$ ($\gg100$; see the following section). {Adopting a higher value for the conversion factor would lower the hydrogen mass estimate.} 

The derived limit then gives {molecular gas-to-stellar mass ratio, $f_{\rm H2}=M_{\rm H2}/M_*<2.3\,\%$ ($3\,\sigma$).} {It is noted, however, that our upper limit of $L_{\rm [CI]}$ locates at the upper bound of the main sequence galaxies presented in \citet{valentino20}. Therefore, the non-detection of [CI] in our data cube is still reasonable for its $L_{\rm IR}$ at the depth of our observations.}

\begin{figure}
\centering
	\includegraphics[width=0.48\textwidth]{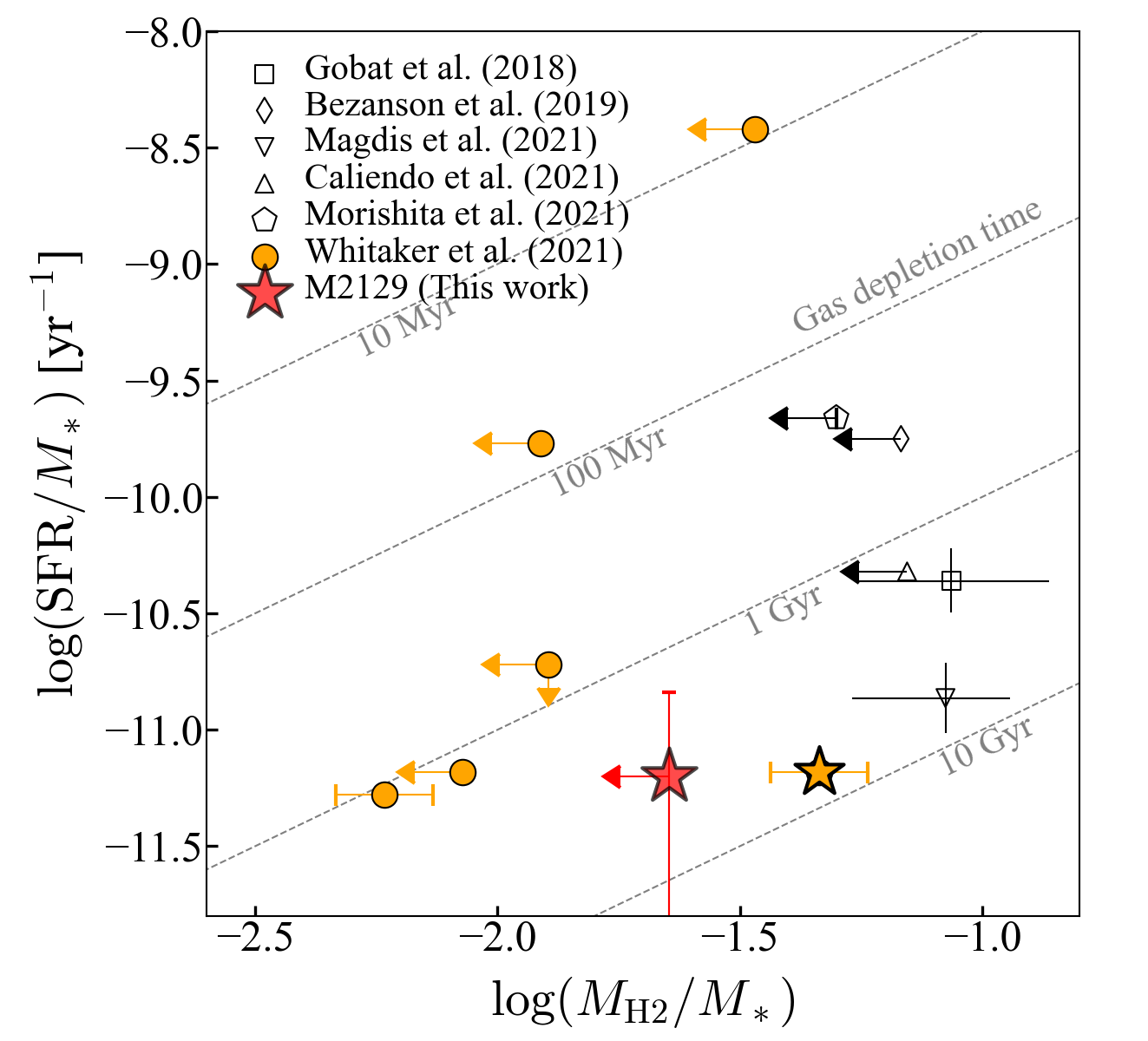}
	\caption{
	Distribution of quiescent galaxies at $z\sim2$ in the molecular gas-to-stellar  ratio ($M_{\rm H2}/M_*$)--specific star formation rate (${\rm SFR}/M_*$) plane. For those undetected, $3\,\sigma$ upper limits of the gas-to-stellar mass ratio are shown. It is noted that two measurements for \targ\ derived in this study (red star; {Table~\ref{tab:sed}}) and \citet[][orange star]{whitaker21} are shown separately. The difference is attributed to two different methods of deriving gas mass (Sec.~\ref{sec:gasmass}).
	}
\label{fig:gfrac}
\end{figure}

In Fig.~\ref{fig:gfrac}, we show the derived upper limit on the gas-to-stellar mass ratio of \targ, along with those of quiescent galaxies in the literature \citep{gobat18,bezanson19,magdis21,caliendo21,whitaker21}. The diagram illustrates the current pace of gas consumption in galaxies, i.e. gas depletion time, defined by $t_{\rm depl}=M_{\rm H2}/{\rm SFR}$. Interestingly, our upper limit on the gas mass fraction of \targ\ is $\sim2\times$ lower than the previous measurement by \citet{whitaker21}, making its gas depletion time decrease by the same factor. The tension is primarily attributed to the assumption on gas-to-dust ratio, where a constant gas-to-dust ratio ($=100$) was adopted in \citet{whitaker21}. A similar value for the ratio was adopted in \citet[][]{gobat18,magdis21,caliendo21}, whereas \citet{bezanson19} derived the upper limit from non-detection of CO(2-1). 

\begin{deluxetable}{lcc}
\tabletypesize{\footnotesize}
\tablecolumns{3}
\tablewidth{0pt} 
\tablecaption{Integrated photometric fluxes of \targ, in units of $\mu$Jy.}
\tablehead{
\colhead{Filter} & \colhead{Wavelength} & \colhead{Flux}\\
\colhead{} & \colhead{$\mu$m} & \colhead{$\mu$Jy}
}
\startdata
HST ACS WFC F435W & 0.43 & $<0.519$\\
HST ACS WFC F475W & 0.48 & $0.627\pm0.309$\\
HST ACS WFC F555W & 0.54 & $0.756\pm0.236$\\
HST ACS WFC F606W & 0.59 & $0.749\pm0.249$\\
HST ACS WFC F625W & 0.63 & $0.895\pm0.382$\\
HST ACS WFC F775W & 0.77 & $1.804\pm0.500$\\
HST ACS WFC F814W & 0.80 & $2.169\pm0.185$\\
HST ACS WFC F850LP & 0.89 & $4.080\pm0.571$\\
HST WFC3 IR F105W & 1.05 & $6.507\pm0.230$\\
HST WFC3 IR F110W & 1.15 & $13.027\pm0.180$\\
HST WFC3 IR F125W & 1.25 & $18.420\pm0.300$\\
HST WFC3 IR F140W & 1.40 & $28.015\pm0.174$\\
HST WFC3 IR F160W & 1.54 & $35.207\pm0.240$\\
HST WFC3 UVIS F225W & 0.24 & $<1.171$\\
HST WFC3 UVIS F275W & 0.27 & $<1.183$\\
HST WFC3 UVIS F336W & 0.34 & $<0.654$\\
HST WFC3 UVIS F390W & 0.39 & $<0.352$\\
SPITZER IRAC CH1 & 3.55 & $87.000\pm16.062$\\
SPITZER IRAC CH2 & 4.51 & $99.900\pm18.434$\\
SPITZER IRAC CH4 & 7.89 & $<39.680$\\
SPITZER MIPS CH1 & 23.44 & $<68.200$\\
ALMA Band6 (2018.1.00035.L) & 1140.48 & $<248.000$\\
ALMA Band6 (This study) & 1199.76 & $191.000\pm55.000$\\
ALMA Band6 (2018.1.00276.S) & 1288.52 & $156.000\pm33.000$\\
ALMA Band3 (2016.1.01591.S) & 2948.53 & $<16.860$
\enddata
\tablecomments{
$2\sigma$ flux errors are quoted for those detected with S/N$>2$. $2\sigma$ upper limits are quoted for the rest. Magnification is not corrected.
}
\label{tab:mag}
\end{deluxetable}

\section{Discussion}\label{sec:disc}

\subsection{What causes the observed low gas-to-dust ratio?}\label{sec:gtd}
We start our discussion on the absence of \ci\ emission in our data. Based on our new ALMA observations and multi-band analyses, we found a significant amount of dust in \targ. The derived $3\,\sigma$ upper limit on the gas-to-dust mass ratio, $\delta_{\rm GDR}\simlt60$, which is, at face value, low for quiescent populations. For example, \citet{whitaker21b} show a wide range of $\delta_{\rm GDR}$, $\sim10$ to $>10000$ of galaxies taken from the SIMBA simulation \citep{dave19}. However, there is a {\it decreasing} trend of \gdr\ with sSFR, such that more quiescence leads to a higher value. With that taken into account, the value found in \targ\ is considered to be relatively low among quiescent galaxies.

Given the observed quiescence of \targ, it is unlikely to expect that dust is decoupled from gas. By following the argument in \citet{mckinnon18}, we obtain the decoupling time $\sim 3\times10^5$\,yr, which is considered to be much smaller than the time scale of its last star formation \citep[][also Sec.~\ref{sec:dust_prop}]{akhshik21}, and thus we consider that gas should be coupled with dust. Therefore, the absence of \ci\ implies that star formation in \targ\ is somehow suppressed while there still exists a gas reservoir. 

One possible explanation for the observed low \gdr\ value is that the gas content is present but in a warm phase, probably at a much higher temperature than the excitation temperature of \ci\ \citep[$\sim 20$-45\,K;][]{walter11,jiao19}. The presence of an active nucleus in its center \citep[][also Sec.~\ref{sec:dust_prop}]{newman18} as a heating source supports this interpretation. It is also noted that the gas mass limit here is derived under the assumption that \ci\ is co-spatial with dust. For example, \citet{gullberg18} found that [C~II] is $\sim1.6\times$ more extended than dust distribution in SMGs.
 
The observed dust temperature ($\sim 27$\,K; Table \ref{tab:fit_global_sed}) can be used as an indicator of the dust-heating radiation. Hirashita \& Chiang (2022, submitted) provide the expected dust temperature under given $\Sigma_\mathrm{SFR}$ (star formation rate surface density) and $\Sigma_\mathrm{dust}$ (dust surface density) with an assumption that the radiation from stars formed in the current episode of star formation is the dominant dust-heating source. Using the (global) SFR and the total dust mass (Table \ref{tab:fit_global_sed}) divided by the surface area, we obtain $\Sigma_\mathrm{SFR}\simeq 3.4\times 10^{-1}~M_\odot~\mathrm{yr}^{-1}~\mathrm{kpc}^{-2}$ and $\Sigma_\mathrm{dust}\sim 1$--$3\times 10^8~M_\odot~\mathrm{kpc}^{-2}$ after correcting for the lensing effect. By using those values, we derive the expected dust temperature of 20\,K.\footnote{We used the RT model in Hirashita \& Chiang (2022, submitted), which takes into account the shielding of dust (i.e.\ multi-temperature effects of dust). If we assume a uniform dust temperature (`one-$T$ model'), the expected dust temperature is even lower.} Note that the local $\Sigma_\mathrm{SFR}$ at the position of the dust cloud is expected to be much lower given the observed color distribution of \targ. This means that the above dust temperature can be regarded as an upper limit expected from the stellar heating.  Therefore, we conclude that the stellar heating alone cannot explain the observed dust temperature in this object. This is consistent with the argument above that an AGN is significantly heating the surrounding materials including the dust cloud. 

{It is noted that the observed dust temperature is, despite the presence of an AGN, rather comparable to those in star-forming galaxies. However, this is not unexpected. For example, \citet{chen21} found no significant difference in dust temperature of AGN host and non-host galaxies at $1<z<3$ \citep[also][]{sinha22}. The similarity in dust temperature of the two populations might be explained by a rapid dust destruction process in the vicinity of an AGN, where the observed dust temperature reflects the one that is not significantly affected (and thus remained) from the radiation.}

\subsection{The presence of an active central black hole and its impact on regulating star formation}\label{sec:dust_prop}
The unprecedented resolution provided by the combination of ALMA and lens magnification enabled us to identify compact dust emission in the center of \targ. The revealed nature of the dust component is of particular interest in the context of massive galaxy quenching at this redshift. 

In Sec.~\ref{sec:gtd} we concluded that \targ\ is likely to harbour an AGN, {as was originally suggested by \citet[][in their Section 7.2]{newman18}. To further investigate the nature of the active central engine, we estimate the bolometric thermal luminosity of the accretion flow toward an AGN.} We start with the \oiii\ line luminosity measured in \citet{newman18}. Dust attenuation is corrected by applying the measured Balmer decrement in the same study (\ha/\hb\,$=3.2$), by following the recipe presented in \citet{lamastra09}. We then convert the \oiii\ luminosity by using an empirical relation presented in \citet{punsly11} and obtain bolometric luminosity $L_{\rm bol}\sim3.2\times10^{45}$\,erg\,/\,s (lens magnification corrected), which is much more powerful than typical Seyfert and in the regime of (type 2) quasars \citep[e.g.,][]{hopkins07,shen20}. {More recently, \citet{man21} revealed a tentative feature of outflows in its Mg~I and Mg~II lines, supporting our interpretation here.}

The presence of an active nucleus is the key to our understanding of possible quenching mechanisms in \targ. Strong outflows from the central AGN seem to be able to explain the observed properties presented here --- spatial offset of dust emission from the stellar center, high dust temperature, and absence of \ci. Since dust grains are easily heated beyond their sublimation temperature and destroyed in the vicinity of a central black hole \citep[e.g.,][]{barvainis87,ishibashi14}, the observed offset of dust may be an apparent effect; it may instead reflect the absence of dust in the very center of \targ. \citet{saito22} also revealed similar features as here in their high-resolution imaging of NGC~1068, a nearby Seyfert galaxy---centrally concentrated dust emission and \ci\ cavity. Through the comparison with several models, they concluded that the observed features are likely due to the interaction of the jet and ionized gas outflow with the galaxy disk originating from negative AGN feedback. While the lack of \ci\ detection in our study prevents us from reaching the same conclusion, the observed properties of \targ\ are suggestive of the presence of strong outflows from the central black hole.

The conclusion here is reached independently from the one in Sec.~\ref{sec:gtd}, which is based on the observed \gdr\ and dust temperature. Those two independent conclusions lead us to the exceptionally unique nature of \targ, that is, the central black hole is still active whereas the overall system is characterized as quiescent. A picture emerges that we may be witnessing the last breath of its central engine in a recently quenched system. 

It is also noted that only a handful of gravitationally lensed type 2 QSOs are known in the literature \citep[e.g.,][]{leung16}. With its strong magnification, \targ\ makes one of the excellent targets that enable in-depth investigation of e.g., co-evolution of supermassive black holes and host galaxies.

\section{Summary}
In this study, we presented new ALMA observations of \targ, a massive quiescent galaxy at $z=2.15$ behind the cluster of galaxies MACS\,2129-0741. With the combination of the angular resolution power of ALMA and the gravitational lensing effect, we revealed the spatial distribution of dust in \targ\ at an unprecedented physical scale, $\sim130$\,pc.

Our data revealed significant detection of very compact continuum emission at $1.2$\,mm, while \ci, as a tracer of molecular hydrogen, was not detected. From this, we placed a $3\,\sigma$ upper limit on the gas-to-dust mass ratio, \gdr\,$<60$, which is $\sim2\times$ smaller than the typical value ($\sim100$) often assumed for quiescent galaxies at $z\sim2$ in the literature. Our finding supports the idea of there being a wide range of \gdr\ in galaxies at this redshift as seen in numerical simulations. This is an important implication, as in the literature often a uniform value is adopted to predict the amount of molecular gas content from dust mass measurements. Follow-up observations of recently quenched galaxies will help us understand if the observed low gas-to-dust ratio is common among those galaxies in the early universe.

Our data revealed its compact dust distribution in a quiescent galaxy at unprecedented angular resolution. Based on the observed properties, supplemented by previous spectroscopic observations, we discussed possible quenching mechanisms that occurred in \targ. The observed properties suggest that outflows from the central active galactic nucleus have played a key role in regulating star formation in \targ. All features characterize \targ\ as an exceptionally unique object, which will allow us in-depth investigation of galaxy quenching and co-evolution of supermassive black holes and host galaxies.

\section*{Acknowledgements}
We would like to thank the anonymous referee for providing constructive comments and suggestions, which improved the manuscript significantly. We are grateful to Katherine Whitaker for kindly providing their flux and dust mass measurements of \targ\ presented in \citet{whitaker21} and for helpful discussion. We are grateful to Dan Coe for his suggestions on our proposal planning. TM is grateful to George Helou, Andreas Faisst, Paul Goudfrooij, Seppo Laine, Matthew Malkan, and Colin Norman for their insightful comments on our findings. We are grateful to Justin Spilker and Ian Smail for their comments on the paper draft.
This paper makes use of the following ALMA data: ADS/JAO.ALMA\#2021.1.00847.S, 2016.1.01591.S, 2018.1.00035.L, and 2018.1.00276.S. ALMA is a partnership of ESO (representing its member states), NSF (USA) and NINS (Japan), together with NRC (Canada), MOST and ASIAA (Taiwan), and KASI (Republic of Korea), in cooperation with the Republic of Chile. 
The Joint ALMA Observatory is operated by ESO, AUI/NRAO and NAOJ. The National Radio Astronomy Observatory is a facility of the National Science Foundation operated under cooperative agreement by Associated Universities, Inc.
Support for this study was provided by NASA through a grant HST-GO-15212 from the Space Telescope Science Institute, which is operated by the Association of Universities for Research in Astronomy, Inc., under NASA contract NAS 5-26555. 
{
{\it Software:} Astropy \citep{astropy:2013, astropy:2018}, numpy \citep{oliphant2006guide,van2011numpy}, \pix\ \citep{abdurrouf21}, python-fsps \citep{foreman14}, EMCEE \citep{foreman13}.
}

\section*{Appendix~A}\label{appen:vmap}
\begin{figure*}
\centering
	\includegraphics[width=0.99\textwidth]{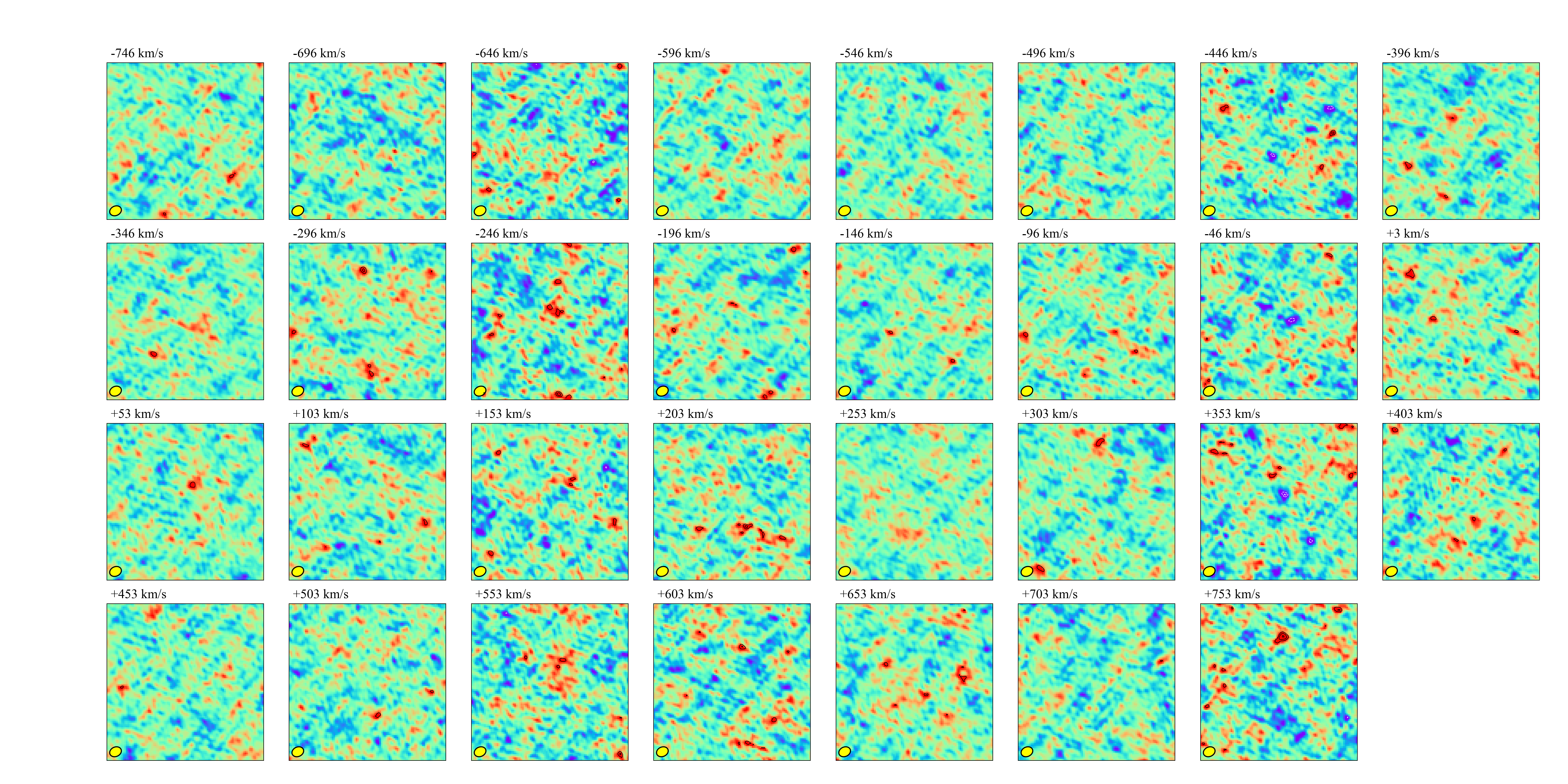}
	\caption{Velocity maps of \ci\ in \targ. The velocity maps are shown at an interval of $\Delta v=50$\,km/s with respect to the \ci\ line at the redshift of \targ. Each stamp has size of $1.\!''5\times1.\!''5$. The contour levels in black solid lines represent positive fluxes at 3\,$\sigma$ and 4\,$\sigma$, in white dashes lines negative fluxes. No confident detection of the \ci\ line is identified.
	}
\label{fig:vmap}
\end{figure*}

In Fig.~\ref{fig:vmap}, we show the velocity map of \targ, around the expected velocity position of \ci. The data cube is generated by running the {\tt tclean} task of CASA on the continuum subtracted visibility data in the way described in Sec.~\ref{sec:alma}, with natural weight, robust parameter of 0.5, and no tapering. While there are positive fluxes identified in some velocity elements (e.g., at $v=246$\,km/s), none of these are detected confidently over more than one velocity element.

\bibliography{sample631}{}
\bibliographystyle{aasjournal}



\end{document}